\documentclass[lettersize,journal]{IEEEtran}
\usepackage{amsmath,amsfonts}
\usepackage{algorithmic}
\usepackage{algorithm}
\usepackage{array}
\usepackage[caption=false,font=normalsize,labelfont=sf,textfont=sf]{subfig}
\usepackage{textcomp}
\usepackage{stfloats}
\usepackage{url}
\usepackage{verbatim}
\usepackage{graphicx}
\usepackage{cite}
\usepackage{subcaption}

\usepackage{tabularx}

\usepackage{color}
\usepackage{color,soul}
\usepackage{multirow}
\usepackage[hidelinks]{hyperref}

\usepackage{lineno}

\begin{document}

\title{Explainable Deep Learning Analysis for Raga Identification in Indian Art Music}

\author{Parampreet Singh, Vipul Arora, ~\IEEEmembership{member,~IEEE,}
\thanks{This paper was produced by the IEEE Publication Technology Group. They are in Piscataway, NJ.}
\thanks{Manuscript received .... .., ....; revised ..... .., ....}}

\markboth{Journal of \LaTeX\ Class Files,~Vol.~...., No.~...., ....~....}
{Shell \MakeLowercase{\textit{et al.}}: A Sample Article Using IEEEtran.cls for IEEE Journals}

\maketitle

\begin{abstract}
Raga identification is an important problem within the domain of Indian Art music, as Ragas are fundamental to its composition and performance, playing a crucial role in music retrieval, preservation, and education.
Few studies that have explored this task employ approaches such as signal processing, Machine Learning (ML), and more recently, Deep Learning (DL) based methods. 
However, a key question remains unanswered in all these works: do these ML/DL methods learn and interpret Ragas in a manner similar to human experts? 
Besides, a significant roadblock in this research is the unavailability of an ample supply of rich, labeled datasets, which drives these ML/DL-based methods.
In this paper, firstly we curate a dataset comprising 191 hours of Hindustani Classical Music (HCM) recordings, annotate it for Raga and tonic labels, and train a CNN-LSTM model for the task of Automatic Raga Identification (ARI).
We achieve a chunk-wise f1-measure of 0.89 for a subset of 12 Raga classes.
Following this, we make one of the first attempts to employ model explainability techniques: SoundLIME and GradCAM++ for Raga identification, to evaluate whether the classifier's predictions align with human understanding of Ragas. 
We compare the generated explanations with human expert annotations and further analyze individual test examples to understand the role of regions highlighted by explanations in making correct or incorrect predictions made by the model. Our results demonstrate a significant alignment of the model's understanding with human understanding, and the thorough analysis validates the effectiveness of our approach.

\end{abstract}


\section{Introduction}
\IEEEPARstart{I}{ndian} Art Music (IAM) is a culturally rich and complex musical tradition that has evolved over centuries. It is categorized into two main categories: Hindustani Classical Music (HCM) and Carnatic music (CM). 
Both forms of music have their own distinct nature and way of singing. Central to IAM is the concept of a Raga, which can be termed as melodic structures that provide a framework for improvisation and composition, making it fundamental descriptor for IAM. It represents the mood of any musical composition. 
Ragas also form the basis of a lot of indian film or folk music\cite{phrase_based_raga}. Detailed description about IAM Ragas can be found at \cite{bor_1995_raga_basics_2,Clayton_1998_raga_basics}.

Comparing with Western Music (WM), a Raga in IAM is more complex than a mode in WM, as it encompasses a broader range of musical features, including a specific scale, characteristic note sequences, hierarchical relationships among notes, relative durations, microtonal intonation, ornamentation (gamaka), melodic movements, specific phrases (pakad), and the evocation of particular moods or emotions\cite{multimodal_HCM,bor_1995_raga_basics_2}.
In WM, we see more emphasis on harmony as compared to HCM, which focuses on melody\cite{Agarwal2013ACS_comparitive_western_HCM}
with improvisation within the framework of ragas.
The task of Raga identification in IAM differs significantly from key or mode detection in WM, as key detection involves identifying the tonal center and scale of a piece\cite{key_detection_tonals_MIT_press_book}, whereas Raga identification requires recognizing complex melodic structures, ornamentations, and expressive nuances that are unique to each Raga.
Apart from that, categories such as Dhrupad, Khayal, Thumri, Dadra, Tappa, and Bhajan, etc, in IAM are considered distinct genres \cite{Agarwal2013ACS_comparitive_western_HCM}, representing different styles or forms of musical performance, and ragas serve as the foundational melodic frameworks within these genres.

For data-driven methods used in computational musicology and music information retrieval (MIR), digital music artifacts with metadata and expert annotations are a prerequisite\cite{Serra_computational_2017}.
The progress of research in IAM is often hindered by the scarcity of a variety of open, labeled datasets.
The introduction of new datasets in IAM  can be useful for making tasks like melody extraction, ARI, tala identification, source separation, automatic mixing, and performance analysis more robust and generalisable.

For the task of Automatic Raga Identification (ARI), various Machine Learning (ML) / Deep Learning (DL) based approaches are being used, but we have no way of knowing if the models are making predictions as if an IAM expert would do or not. This motivates us to look for these answers from the field of Explainable AI (XAI), which has barely been explored in the field of IAM analysis.
XAI is a popular research area within ML that aims to provide explanations about complex deep learning models, making them interpretable \cite{gradcam,gradcam++,saumitra_mishra_2018_soundlime}.
XAI focuses on explaining the behavior of models and their predictions in various known and unknown situations in a post-hoc manner, hence increasing transparency and understanding of these models, ultimately improving their reliability and trustworthiness. 
In this study, we take up the task of ARI for IAM and then use XAI methods to show the reliability of the classifier.

The key contributions of this work are as follows:
\begin{itemize}
\item We manually annotate an HCM dataset using audio recordings from Prasar Bharti. 
The sample audios, along with metadata information, can be accessed at: \url{https://github.com/ParampreetSingh97/PIM_v1_ExAI.git}.
\item We train a CNN-LSTM-based model for the task of ARI and benchmark its performance on the annotated dataset. 
\item We make one of the first attempts at using XAI techniques for ARI and subsequently validating the significance and correctness of the Raga classifier musically.
\item We show that the ARI model in fact does make musically correct predictions, making it usable for various HCM tasks like music teaching, automatic labeling, music suggestion and search, and other tasks involving IAM Ragas.
\end{itemize}

\section{Related Work}\label{sec:Related Work}
\subsection{HCM datasets}
The Indian Art Music Raga Recognition Dataset (RRD)\cite{raga_recognition_dataset}, offers 116 hours of HCM with 300 recordings featuring 30 unique Ragas, with 10 recordings per Raga.  
Then there is Saraga dataset \cite{Saraga}, which comprises of 108 recordings of HCM in 61 unique Ragas and 9 unique talas, totaling 43.6 hours. 
Both of these datasets feature live performances by various experts, along with providing labels like Raga names, manually marked melodic phrases, and features like tonic and pitch extracted using various algorithms. 
While Saraga allows download access with annotations, audio files RRD are currently unavailable.
Notably, we observe a significant overlap of audio files between Saraga and RRD.
Some other works \cite{bidkar12north,ml_classifiers_for_raga_recognition} 
utilize self-made HCM datasets for ARI, but these datasets have not been made publicly available.

Our dataset stands out as the  largest labeled HCM dataset to the best of our knowledge, surpassing both Saraga and RRD, in terms of total duration, number of recordings, unique Ragas, talas and manually labeled Raga and tonic information.
\subsection{Raga Identification}
For HCM, the paper \cite{bidkar12north} uses their own generated instrumental dataset of 20 second snippets for 12 Raga classification. They make use of a combination of features like Pitch, MFCC and other statistics derived from mfcc together for the classification task.
Another paper \cite{multimodal_HCM} does 2-Raga classification using a subset from Saraga dataset and report that using Swargram features is helpful for Raga classification, which is nothing but chromagram features shifted in pitch.
The paper \cite{Melodic_pattern_recog_2018_HCM} uses pitch class distribution and N-gram approach on the extracted pitch for ARI and reports that longer audio clip length is beneficial for the task. 
Another work \cite{priya} uses a dataset downloaded from YouTube for the task of 6-class classification.
In paper~\cite{raga_using_ML}, KNN and SVM machine learning algorithms were used for classifying different Ragas like Yaman and Bhairavi, and the KNN was found to be performing slightly better. 
The paper~\cite{phononet} proposed a hierarchical deep learning system called PhonoNet, which used a CNN model in the first stage to predict the Raga of a small audio chunk, and a two-stage recurrent LSTM network in the second stage to look at various parts of the audio input and prioritize observed patterns.

There are some other works related to CM\cite{ phrase_based_raga,modeling_and_analysis_ieee,raga_using_cnn,carnatic_deeplearning,scale,cla,raga_survey,acous}, and most of them use basic features like extracted pitch, MFCCs, spectral centroid, chromagram, spectrogram, scalogram etc and apply different ML classifiers like Naive Bayes, SVM, Random forest, GMMs, and some also use DL models like RNN, LSTM, CNN etc. 

 We observe that we have more research centered around CM, primarily because of higher availability of labeled datasets as compared to HCM. Also, to the best of our knowledge, none of the works in the literature have explored XAI methods for the task of ARI in any form. 
\subsection{XAI methods}
Until now, we have a lot of research already available focused on XAI methods for DL models used for various tasks involving images, and some works for audio and music also. 
One very prominent kind of approach for post-hoc explainability involves the use of model's gradients from various layers of a CNN or ANN model and using them in an intelligent way to visualize explanations of the classifier in the form of saliency map superimposed over the original image itself, highlighting the most important region for classification by the classifier \cite{Vanilla,Guided_backprop,smoothgrad,cam,gradcam,gradcam++,LRP,Integrated_gradients}. We will be using GradCAM++\cite{gradcam++}, which is a very popular method for CNN explanations. It weighs each of the pixels in each feature map of the last CNN layer separately, with weights determined using the gradients of these layers, 
and then finally stacking all the weighted feature maps together to get the final attribution map, which highlights important regions in an image. 
In the audio domain, we have some works\cite{deepfake} which treat audio feature representations as images and use the XAI methods for images to generate explanations.

Other type is of black box models, whose gradients are not accessible and we have to design XAI methods by querying the model multiple times. There are some notable works for black box model explainability, with LIME\cite{lime} generating locally faithful explanations by training an interpretable model on perturbed instances of the input data sample.  
Another method SHAP\cite{lundberg2017unified_shap}, uses game theory to assign a unique contribution score to each feature in the input, indicating its impact on the model's output. SHAP is computationally intensive, especially for models with a large number of features. 
Another approach Anchors\cite{Ribeiro2018Anchors} express explanations as simple IF-THEN rules, making them more easily understandable, but it struggles with large and complex datasets. 
  
For audio, we have a variant of LIME in SoundLIME\cite{saumitra_mishra_2018_soundlime}, which works on spectrograms or other audio-based features, treats them as image data only and then replicates the LIME framework. We use SoundLIME out of the black box models for our analysis.
We employ both white-box and black-box XAI models to ensure comprehensive coverage of interpretability techniques, enabling a thorough comparison and analysis of the performance of our ARI model.

\section{Dataset}\label{sec:Dataset}
\begin{figure}[!t]
\centering
\includegraphics[width=1\columnwidth]{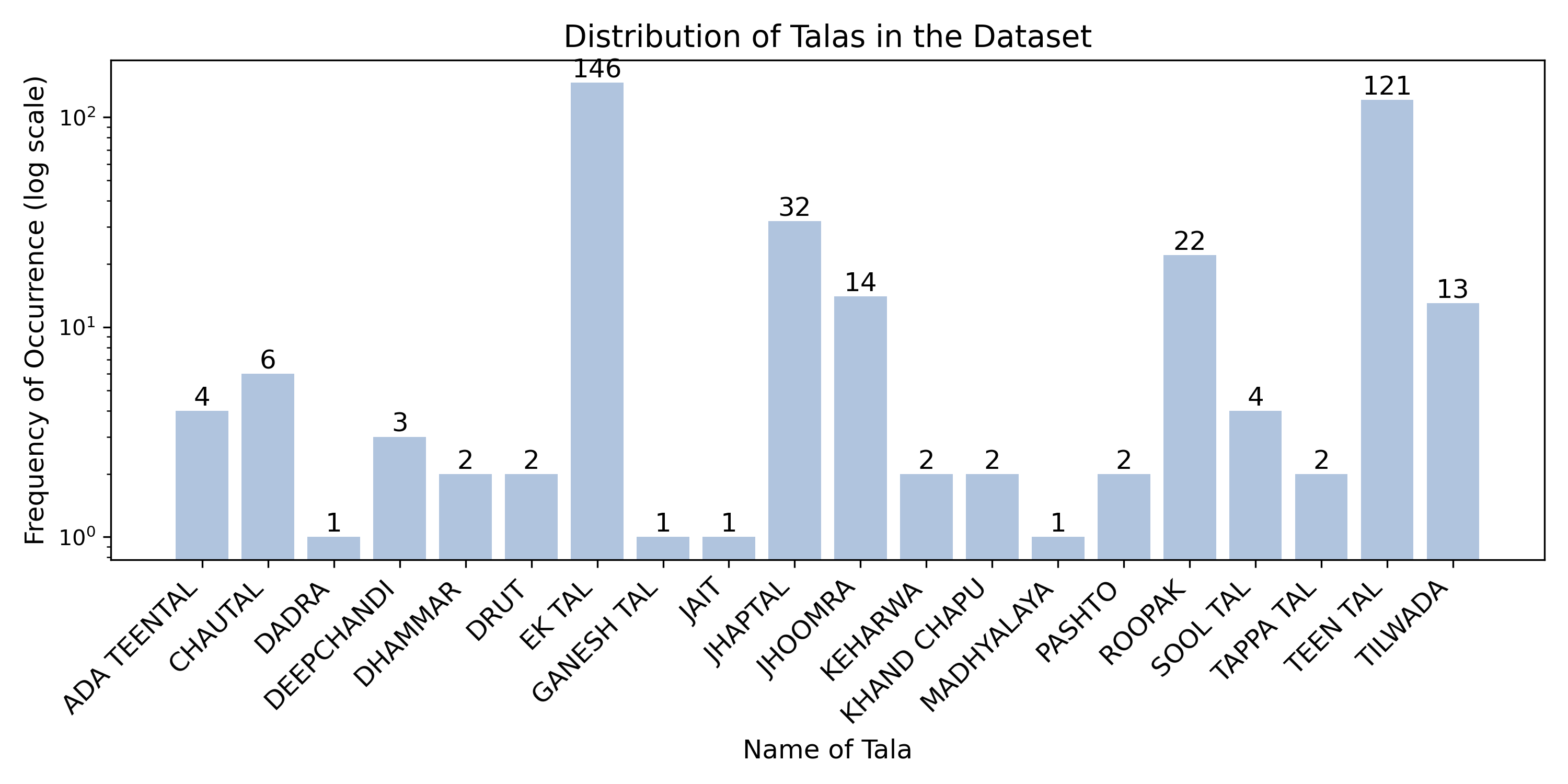}
\caption{Distribution of number of audio files per tala}
\label{fig_talas}
\end{figure}

The dataset comprises a total of 501 HCM songs, with each song being a polyphonic audio containing vocals of the main artist, percussion instruments like tabla, Pakhawaj, along with Manjeera in some audios, along with non-percussion instruments like Harmonium, Sarangi, Violin.
These music recordings are sung in variety of Ragas and Talas with all of the songs being of varying lengths. The dataset includes 144 unique Ragas performed by 169 vocal artists. 
Harmonium is the most common instrument, present in a total of 372 songs, while some performances feature multiple string instruments. All of the songs include drone (tanpura) playing in the background of the performance, which provides the tonic for the whole music performance. The dataset includes performances in 20 different talas accompnying the vocals, with teentaal (121 instances) and ektaal (146 instances) being the most frequently used talas as shown in Fig.~\ref{fig_talas}.
Each song is available in the form of 30-second chunks, totaling 23,005 audio files, approximately 191.70 hours of music.

\subsection{Musical Styles}
The dataset covers a variety of musical styles belonging to HCM, such as Khayal, Dhrupad, Dadra, Thumari, Tarana, Gat, Bhajan, Tappa, Dhammar, Kajri, and Jhoola (folk), showcasing a wide range of songs from different regions throughout India.

\subsection{Recording Source and Quality}
All recordings are in the form of mono tracks, sourced from the archives of Prasar Bharti (Akashwani, Delhi), which are not available online elsewhere. They feature performances by HCM ustaads (experts) and are mainly studio recordings or recordings from live shows, ensuring high audio quality. Most of the featured vocal, percussion, or string instrument artists are well-known in the HCM community. 
Some of the recordings feature speech at the start which includes introduction of the artists along with details about their performance including the Raga they will be performing. The speech segments range from a few seconds to about a minute or two in some recordings.

\subsection{Metadata}
The metadata contains details such as the names of vocal and instrumental artists, the instruments played, the event where the performance took place, the name of the song, and the number of 30-second audio files for each song. We also add to the metadata the Raga name, the tonic and the start and end times of music and speech for each of those songs, labeled manually by our annotators.

\subsection{Manual Annotation Process} 
The labeling process for the Raga and Tonic labels is carried out by two expert vocalists, each with over 10 years of experience in HCM. The experts are provided with 501 complete audio files, each containing full performances, to label Raga names and tonic values. These labels are then applied to all the 30-second chunks derived from each file.

In many of these audio files, there is a narrator who announces the raga to be performed, ensuring the correctness of the raga label. Additionally, certain files come with pre-assigned Raga labels from Prasar Bharti. For these files, our experts review and verify the pre-assigned labels, half files each. For the remaining files without pre-assigned labels, the two experts independently label half of the files each. To further ensure consistency in labeling and minimize subjectivity, a third expert reviews all the Raga and Tonic labels throughout the dataset.

\section{Methodology}\label{sec:Methodology}

In this study, we use our curated dataset to extract chromagram features from the audios and train a CNN-LSTM model to perform multi-class classification for $N=12$ Raga classes, which represent the top classes with the most audio files in our dataset, ensuring sufficient examples for splitting the data into training, validation, and test sets. By limiting $N$ to just 12 classes, we aim to create a high performing classifier model, which is crucial for producing reliable results in our XAI analysis.
Next, we employ Grad-CAM++ (GC) and SoundLIME (SL), two popular XAI models, for saliency visualization and compare them with annotations from music experts.
We also train the same model on the whole dataset, but we do not use it for XAI analysis because of performance shortcomings.
In this section, we explain the entire procedure. 

\subsection{Feature Extraction}\label{sec:Feature_extraction}
For each 30-second audio chunk, we extract pitch-based chromagram features using the librosa library in Python. These features are computed with $n_{chroma}$ chromagram bins. The chromagram provides a time-frequency representation emphasizing the pitch classes, which is critical for raga analysis.
However, before inputting the chromagram to the Deep Learning model, the signal undergoes tonic normalization as explained.

\subsection{Tonic Normalisation}\label{sec:Tonic Normalisation}
Our dataset contains performances by various artists, with each performance having a single tonic note that serves as the reference note for constructing the Raga. However, different performances in the same Raga have been sung in varying tonics, leading to structural differences even within the same Raga. To address this, we perform tonic normalization on the extracted chromagram features, aligning all performances to a common reference note. 
 
For a given audio signal $x(t)$, we first compute its Short-Time Fourier Transform (STFT), denoted as $X(n, k)$. The magnitude squared of the STFT coefficients $|X(n, k)|^2$ yields the spectrogram of the signal, capturing the energy distribution over time and frequency.
To obtain a pitch-based log-frequency spectrogram, we map the STFT frequency bins to the 128 MIDI pitch values. The center frequency of each MIDI pitch $p$ is given by \cite{muller}:
\begin{equation}
\label{eqn:eq_f_pitch} 
F_{\text{pitch}}(p) = 440 \times 2^{\frac{(p - 69)}{12}}, \quad \text{for } p \in [0, 127]
\end{equation}
For each pitch \( p \), we define a set \( P(p) \) as:
\begin{equation}
\label{eqn:pp_eqn} 
P(p) = \left\{ k \mid F_{\text{pitch}}(p - 0.5) \leq F_{\text{coef}}(k) < F_{\text{pitch}}(p + 0.5) \right\}
\end{equation}
This effectively groups the STFT frequency bins into 128 pitch bands corresponding to the MIDI pitches.
Now, we have the log-frequency spectrogram \( Y_{\text{LF}}(n, p) \) given by:
\begin{equation}
\label{eqn:lf_spect} 
Y_{\text{LF}}(n, p) = \sum_{k \in P(p)} |X(n, k)|^2
\end{equation}
This reduces the frequency axis of the original spectrogram (which has $ (N_{\text{FFT}}/2) + 1 $ frequency bins) to 128 values, each corresponding to a MIDI pitch. 
To obtain chromagram features, we simply sum up the pitch values which differ by some integer multiple of an octave as they are perceived to be the same note musically\cite{muller}:
\begin{equation} 
\label{eqn:chroma} 
C(n,c)  = \sum_{p\in[0:127]:p mod 12=c} Y_{LF}(n,p)
\end{equation} 
Now, to shift the chromagram representation of any song by a factor of $``a"$ semitones, we are essentially looking to get a new $X'(n,k)$ such that:
\begin{equation} 
\label{eqn:chroma_shift} 
X'(n,k) = X(n,k\times2^{a/12})
\end{equation} 
Shifting by $``a"$ semitones essentially means multiplying the frequency values in a spectrogram by $2^{a/12}$, where $``a"$ is a discrete value such that $a\in[1:11]$. 
We choose \( a \) within this range because shifting by 12 semitones (an octave) would map notes back to their original pitch classes, which is not desired for this transformation.
Now, we will get the log-frequency spectrogram as:
\begin{equation} 
\label{eqn:lf_spect2} 
Y_{LF}'(n,p)=\sum_{k\in P(p)} |X(n,k')|^2
\end{equation}
where, $k'= k\times2^{a/12}$ \\
Now, using Equation: \ref{eqn:eq_f_pitch},
\begin{equation} 
\label{eqn:f_pitch_2} 
F_{pitch}(p) = k' = 440 \times 2^{(p-69)/12}
\end{equation}
Substituting the value for $k'$, and re-arranging terms, we get:
\begin{align}
k &= 440 \times 2^{((p-a)-69)/12} \label{eqn:k_shift}
\end{align}
Equation~\ref{eqn:k_shift} shows that shifting the frequency bins by \( a \) semitones corresponds to shifting the MIDI pitch numbers by \( a \) steps downward. Thus, each pitch \( p \) in the shifted spectrogram \( Y_{\text{LF}}'\left( n, p \right) \) maps to pitch \( p - a \) in the original spectrogram \( Y_{\text{LF}}\left( n, p \right) \).
This implies that shifting the chromagram by \( a \) semitones is equivalent to a cyclic shift of the pitch axis by \( a \) steps. Since the chromagram representation has a periodicity of 12 semitones (an octave), shifting by \( a \) semitones wraps the pitch classes cyclically.
So, to apply tonic normalization to our audio data, we can simply perform a cyclic shift of the extracted chromagram bins along the pitch axis by \( a \) steps.
\begin{figure*}[!t]
    \centering
    \includegraphics[width=1\textwidth]{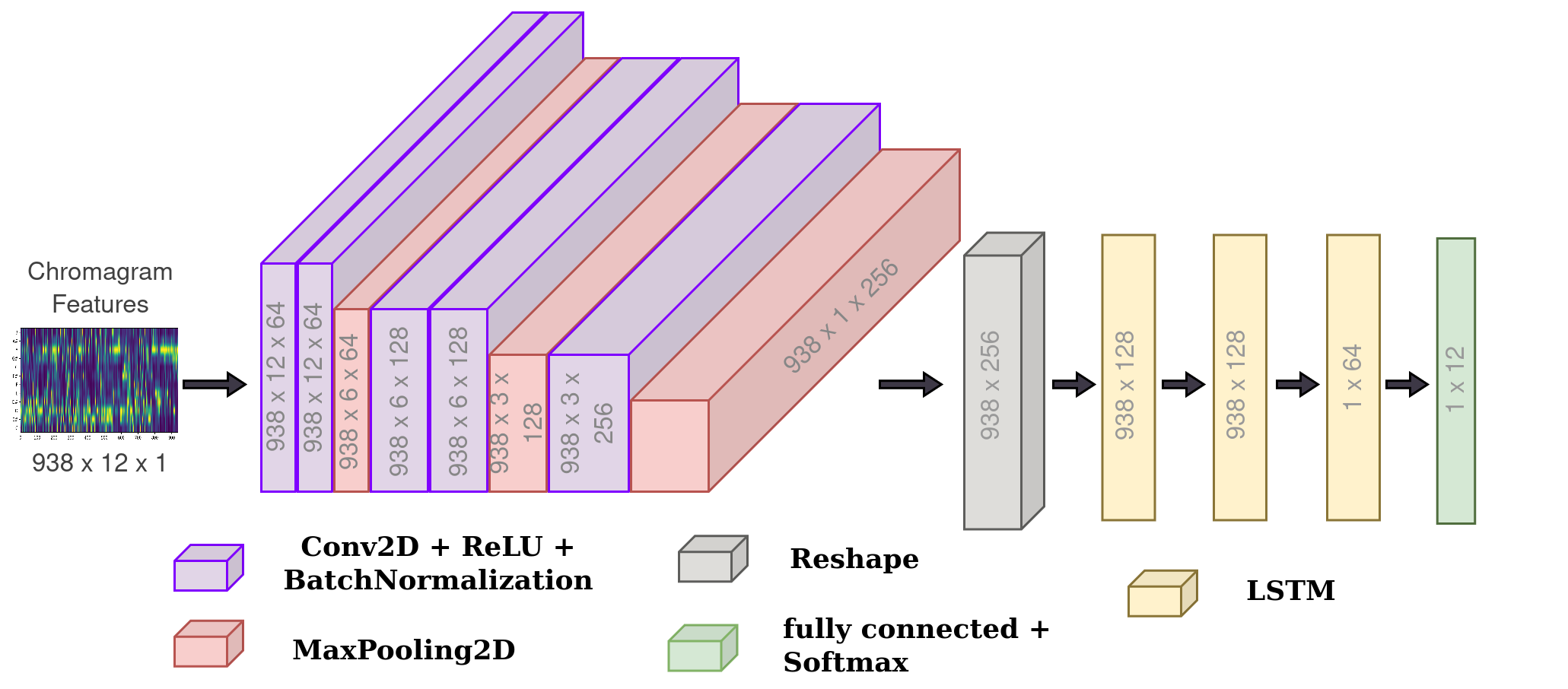}
    \caption{Model Architecture for CN2+LSTM+T}
    \label{fig:Model_arch}
\end{figure*}
\subsection{Classification Model Description}\label{sec:Model Description} 
To determine the best-performing model, we conduct an ablation study, experimenting with various CNN and LSTM-based architectures. Our first model, CN1+T, utilizes a simple CNN architecture with convolutional, max-pooling, and batch normalization layers, followed by a flatten layer, in a very similar way to a common image classification model. Then the CN1+LSTM+T model extends this by replacing the flatten layer with an LSTM layer, ideal for processing time-series data like audio signals. Also, we provide tonic normalization as discussed in Section:\ref{sec:Tonic Normalisation} and test the model performance. 
The CN2+LSTM model features a more complex CNN-LSTM architecture with 1D max-pooling to preserve temporal information, as shown in Fig.~\ref{fig:Model_arch}. Then the CN2+LSTM+T model just includes tonic normalization in the input features, following the exact configuration of CN2+LSTM model. The selection of layer configurations and activation functions is carefully optimized for classification performance. These models will be further detailed in section\ref{sec:Results}, where we evaluate their effectiveness in ARI.

\subsection{Classification Model Training}\label{sec:Model Training}

We train our DL models for the 12-class classification task using softmax activation in the last layer. Let \( z_i \) be the output logit for the \( i^{\text{th}} \) neuron in the last dense layer. The softmax function converts these logits into probability values:
\begin{equation}
\label{eqn:softmax}
p_i = \text{softmax}(z_i) = \frac{e^{z_i}}{\sum_{j=1}^{N} e^{z_j}}
\end{equation}
where \( N = 12 \) is the number of classes. The probabilities \( p_i \) satisfy \( \sum_{i=1}^{N} p_i = 1 \). We use the categorical cross-entropy loss function for training:
\begin{equation}
\label{eqn:Loss}
L_{\text{CCE}} = -\sum_{i=1}^{N} t_i \log(p_i)
\end{equation}
where \( t_i \) is the true label encoded as a one-hot vector (i.e., \( t_i = 1 \) for the correct class and \( 0 \) elsewhere). Training is performed over multiple epochs until convergence, and model performance is evaluated using precision, recall, and F1-score. We employ cross-validation to assess the robustness and generalization ability of our model.
\subsection{Grad-CAM++}\label{sec:Grad-CAM++}
We utilize the CNN component of our trained model without the LSTM layers to get 256 learned feature maps in our case, as is visible in Fig.~\ref{fig:Model_arch}. To create a class-specific GradCAM++ activation map, each of these feature maps is weighted, and then a weighted sum is taken to generate the final activation map or saliency map for the target class. We compute the weights of feature maps using the weights of those last CNN blocks, given by the equation:
\begin{equation} 
\label{eqn:GC_weights} 
\alpha_c^k = \frac{1}{Z}\sum_i \sum_j \frac{\partial y^c}{\partial A^k_{ij}} 
\end{equation} 
Here, \( A_{ij}^k \) represents the activation at position \( (i, j) \) in the \( k^\text{th} \) feature map. We have \( k \in [1, 256] \) feature maps, and \( c \in [1, 12] \) classes.
$\alpha_c^k$ represents the weights of $k^{th}$ feature map concerning the $c^{th}$ class. $Z$ is the dimension of each feature map, and $y^c$ represents the output logits for class c. 
By computing these weights $\alpha_c^k$, we perform a weighted sum of the feature maps to obtain the GradCAM \cite{gradcam} attribution map given by:
\begin{equation} 
\label{eqn:GC_eq} 
L^c_{Grad-CAM}= ReLU\Bigl(\sum_k \alpha_k^c A^k\Bigl) 
\end{equation} 
After that, ReLU is applied to keep only the positively impacting salient regions and remove the regions with negative gradients in the output saliency map or attribution map. 
Now, we weigh each pixel separately rather than weighting the whole feature map with a single weight. The new weights are given by: 
\begin{equation} 
\label{eqn:GC+_weights} 
\alpha_c^k = \sum_i \sum_j w^{kc}_{ij} ReLU \Bigl(\frac{\partial y^c}{\partial A^k_{ij}}\Bigl) 
\end{equation} 
Here, $A^k_{ij}$ represents each pixel in the $k^{th}$ feature map. This $w^{kc}_{ij}$ aims to weigh each pixel separately rather than the whole activation map. So, this $w^{kc}_{ij}$ is given by: 
\begin{equation} 
\label{eqn:GC+_alpha} 
w^{kc}_{ij} = \frac{\frac{\partial y^c}{(\partial A^k_{ij})^2}}{2\frac{\partial y^c}{(\partial A^k_{ij})^2} + \sum_a \sum_b A^k_{ab} \Bigl( \frac{\partial^3 Y^c}{(\partial A^k_{ij})^3}\Bigl)} 
\end{equation} 
Using these, we finally obtain the attribution map for Grad-CAM++ \cite{gradcam++} given by: 
\begin{equation} 
\label{eqn:GC+_eq} 
L^c_{Grad-CAM++}= ReLU\Bigl(\sum_k \alpha_k^c A^k\Bigl) 
\end{equation} 
This activation map given by $L^c_{Grad-CAM++}$ represents the desired visualisations for the regions of interest in a particular test example for a given class for the classifier model.
\subsection{SoundLIME} SoundLIME\cite{saumitra_mishra_2018_soundlime} is specifically designed for audio, analogous to LIME's\cite{lime} application in images. Here also, we need the pre-trained classifier model, but we treat it as a black-box model with no access to its gradients. We temporally divide the chromagram into 30 equal parts or super-pixels, with one super-pixel representing the chromagram representation for each second of the audio. Next, we create 150 random perturbations for the given test input by applying random multi-hot masks of shape 30x1, meaning that each element of the mask either suppresses or retains each super-pixel. An exponential kernel is defined on L2 distance function $D(x,z)$ between the original image $x$ and the perturbed image $z$, denoted by $\pi_x(z)$:
\begin{equation} 
\label{eqn:kernel} 
\pi_x(z) = exp(-D(x,z)^2/\sigma^2)
\end{equation}
We train a weighted linear regression model, using the mask as input and querying the original classifier model for prediction of the masked input image of the chromagram, using that as the target. The significance of this approach can be understood from the fact that if most of the salient regions are already present in the masked input example and only non-salient regions are masked, the classifier model predicts the correct class. That becomes the label for our linear regression model that whether the given perturbed example results in the desired prediction or not. So, our weighted linear regression model trained on these 150 perturbed data points learns the coefficients for each dimension of the mask and ultimately outputs which regions (super-pixels) of the mask are more salient for the prediction of the current test example.

Let $g \in G$ represents a family of interpretable models, in our case, $g$ represents the linear regression model. $f$ is the classifier model. $\omega(g)$ represents the complexity of the model, in our case it could be represented by the number of non-zero weights. The LIME explanation can be shown mathematically as:
\begin{equation} 
\label{eqn:SoundLime} 
\zeta(x)= argmin_{g\in G}L(f,g,\pi_x) +\omega(g)
\end{equation} 

\subsection{Saliency Evaluation Method}\label{sec:Saliency Evaluation Method} 
To evaluate the salient regions identified by the two XAI models, we use the output of XAI models and compare them with the expert-annotated salient regions. 
We compute the precision for top 1 to 10 seconds for both the XAI models. In our task, the concern lies solely with determining whether the top predicted regions are correct. Hence, precision is the best metric to employ in this situation, whereas recall and accuracy are not required. We then analyse one correct and one incorrect prediction of the classifier for their salient region. 

\section{Experiments and Results}\label{sec:Results}

\subsection{Data Pre-processing}\label{sec:Data Pre-processing} 
\begin{figure}[!t]
\centering
\includegraphics[width=1\columnwidth]{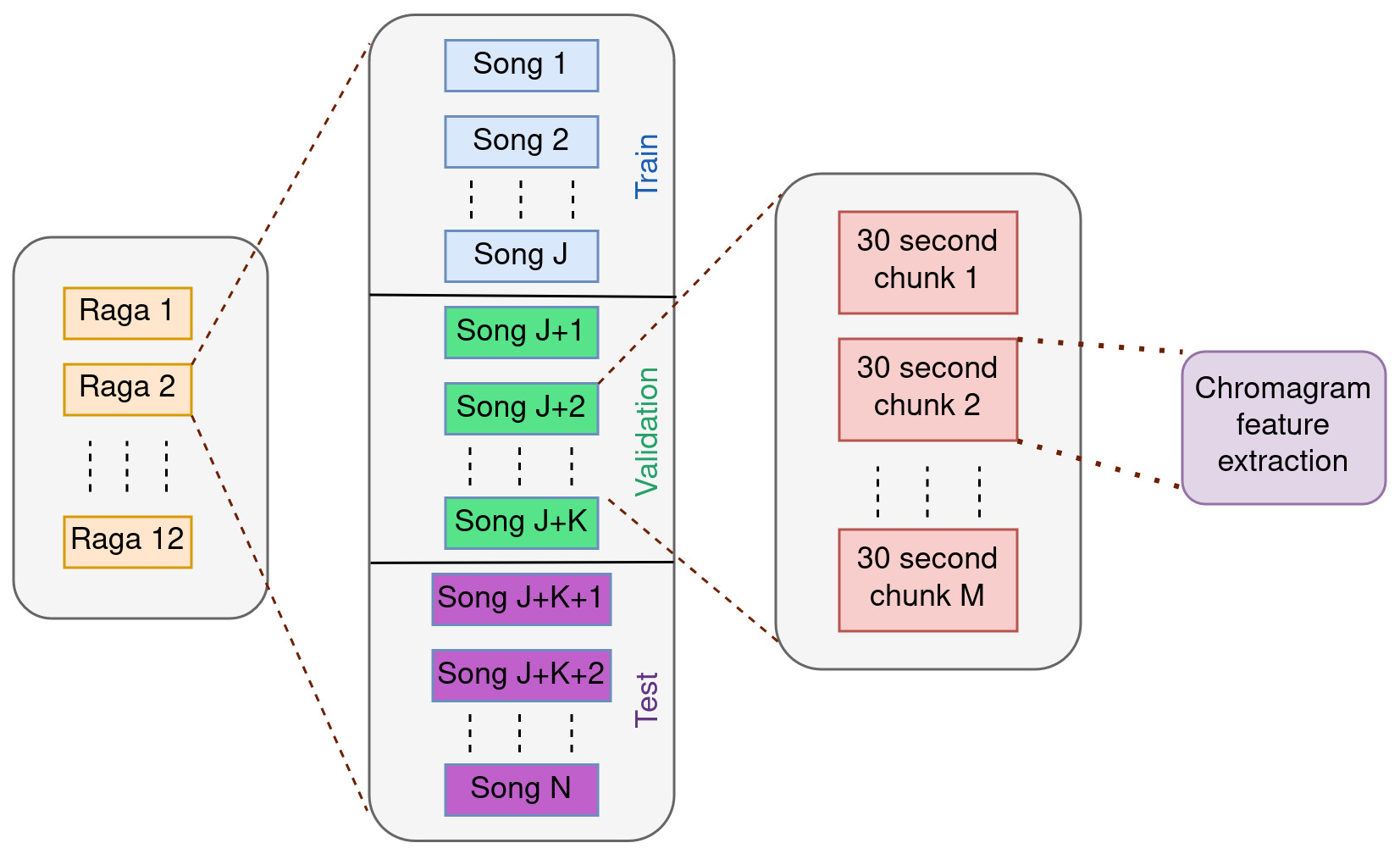}
\caption{Dataset pre-processing for the network}
\label{fig_1}
\end{figure}

To prepare the audio files for analysis, all non-musical speech is removed. 
We distribute all the songs into train-set $S_{tr}$, test-set $S_{t}$, and validation-set $S_{val}$ with percentages of approximately 79\%, 15\%, and 6\% respectively, making sure that there is at least 1 song of each class in the test and validation sets. Then we split the songs into 30-second chunks for all the 3 splits as shown in Fig.~\ref{fig_1}.
Notice that we first split the dataset into training, validation, and test sets at the song level, and then divide audios in each set into 30-second chunks. This ensures that the chunks from the same song are not distributed across different sets, which would introduce bias and compromise the generalizability of our model.
The last chunk of each music recording is purposely removed as it is mostly found to have only 3 to 10 seconds of music followed by silence, or may simply consist of a single ending note prolonged to a few seconds, which may not even contain any salient region for getting classified to a particular Raga.

We extract chromagram features for these audio snippets, making each input sample of shape (938,12), having 938 time frames with 12 chromagram bins each.
We decide Note ``A" to be the reference note for the dataset and carry out
tonic normalization as explained in Section~\ref{sec:Tonic Normalisation}, using the tonic values labeled by the music experts.
Now we carry out training using the train-set $S_{tr}$ and validation-set $S_{val}$ as explained in Section~\ref{sec:Model Description},\ref{sec:Model Training} and then evaluate on test-set $S_{t}$.

\subsection{Classification results}\label{sec:Classification results} 
Chunk-wise classification results on $S_{t}$ after carrying out ablation studies, and cross-validation are shown in Table~\ref{tab:classification_results_combined} (all values rounded off to 2 decimal places) 
\begin{table*}[!t]
    \centering
    \caption{\label{tab:classification_results_combined}Combined Classification Report for Different Models}
    \begin{tabular}{|l|l|l|l|l|l|l|l|l|l|l|l|l|}
    \hline
    \multicolumn{13}{|c|}{\textbf{Classification Report for Different Models}} \\
    \hline
    \textbf{Raga} & \multicolumn{3}{|c|}{\textbf{CN1+T}} & \multicolumn{3}{|c|}{\textbf{CN1+LSTM+T}} & \multicolumn{3}{|c|}{\textbf{CN2+LSTM}} & \multicolumn{3}{|c|}{\textbf{CN2+LSTM+T}} \\
    \hline
    & \textbf{Precision} & \textbf{Recall} & \textbf{F1} & \textbf{Precision} & \textbf{Recall} & \textbf{F1} & \textbf{Precision} & \textbf{Recall} & \textbf{F1} & \textbf{Precision} & \textbf{Recall} & \textbf{F1} \\
    \hline
    \textbf{Bhairavi} & 0.89 & 0.53 & 0.67 & 0.83 & 0.74 & 0.78 & 0.0 & 0.0 & 0.0 & 0.93 & 0.91 & \textbf{0.92} \\
    \hline
    \textbf{Bihag} & 0.42 & 0.31 & 0.36 & 0.87 & 0.77 & 0.81 & 0.40 & 0.74 & 0.52 & 0.90 & 0.91 & \textbf{0.90} \\
    \hline
    \textbf{Des} & 0.09 & 0.08 & 0.08 & 0.63 & 0.74 & 0.68 & 0.06 & 0.08 & 0.07 & 0.78 & 0.66 & \textbf{0.70} \\
    \hline
    \textbf{Jog} & 0.85 & 0.95 & 0.90 & 0.97 & 0.86 & \textbf{0.91} & 0.88 & 0.88 & 0.88 & 0.91 & 0.89 & 0.90 \\
    \hline
    \textbf{Kedar} & 0.72 & 0.81 & 0.76 & 0.92 & 0.75 & 0.82 & 0.0 & 0.0 & 0.0 & 0.93 & 0.89 & \textbf{0.91} \\
    \hline
    \textbf{Khamaj} & 0.24 & 0.38 & 0.29 & 0.70 & 0.85 & \textbf{0.77} & 0.0 & 0.0 & 0.0 & 0.55 & 0.68 & 0.59 \\
    \hline
    \textbf{Malkauns} & 0.94 & 0.89 & 0.92 & 0.90 & 0.96 & 0.93 & 0.64 & 0.61 & 0.62 & 0.99 & 0.96 & \textbf{0.97} \\
    \hline
    \textbf{Maru-Bihag} & 0.57 & 0.56 & 0.56 & 0.80 & 0.90 & 0.84 & 0.52 & 0.68 & 0.60 & 0.88 & 0.85 & \textbf{0.86} \\
    \hline
    \textbf{Nayaki-Kanada} & 0.64 & 1.0 & 0.78 & 0.91 & 0.96 & \textbf{0.93} & 0.17 & 0.20 & 0.19 & 0.93 & 0.92 & 0.92 \\
    \hline
    \textbf{Shudha-Kalyan} & 0.95 & 0.37 & 0.53 & 0.79 & 0.75 & 0.77 & 0.67 & 0.67 & 0.67 & 0.89 & 0.94 & \textbf{0.92} \\
    \hline
    \textbf{Sohni} & 0.35 & 0.61 &  0.45 & 0.88 & 0.90 & \textbf{0.88} & 0.0 & 0.0 & 0.0 & 0.89 & 0.83 & 0.85 \\
    \hline
    \textbf{Yaman} & 0.66 & 0.86 & 0.75 & 0.87 & 0.81 & 0.84 & 0.68 & 0.52 & 0.60 & 0.90 & 0.91 & \textbf{0.90} \\
    \hline
    \textbf{Weighted Avg} & 0.68 & 0.64 & 0.63 & 0.85 & 0.83 & 0.83 & 0.50 & 0.50 & 0.50 & 0.90 & 0.89 & \textbf{0.89} \\
    \hline
    \end{tabular}
\end{table*}
\subsubsection{\textbf{CN1+T}} This model achieved an overall weighted f1-score of 0.78 on $S_t$ for  $N=12$ classes.
The maximum and minimum f1-scores during cross-validation were 0.80 and 0.73, respectively.  
Clearly, this architecture is not suitable for music data, as flattening does not appropriately handle the dependent frame-wise information required in case of such time-series data.
\subsubsection{\textbf{CN1+LSTM+T}} This model showed improved classification performance with an f1-score of 0.83 after cross-validation. The maximum and minimum f1-score observed during cross-validation were 0.87 and 0.79, respectively. This model outperformed the CN1+T model, highlighting the importance of incorporating temporal modeling layers, such as LSTM, instead of flattening layers to effectively capture sequential patterns in music-based applications.
\subsubsection{\textbf{CN2+LSTM}} 
Despite multiple training attempts, extensive hyperparameter tuning, and architectures similar to our main model (CN2+LSTM+T) or its deeper variants with additional CNN and LSTM layers, this model failed to achieve satisfactory performance even on $S_{tr}$. As a result, its performance on the $S_t$ was expectedly poor, underscoring the importance of tonic normalization for the ARI task.
We also experiment with simplifying the problem by carrying out training on a smaller subset of $S_{tr}$ having $N<12$ classes, removing the classes with very low f1-scores during training. While this led to some improvement in training, it suggests a lack of generalization to any number of Raga classes using this approach.
\subsubsection{\textbf{CN2+LSTM+T}} 
This model consistently achieved the highest performance among all models, with an average f1-score of 0.89 after 6-fold cross-validation. The maximum and minimum f1-scores observed during cross-validation are 0.94 and 0.84, respectively. While the CN1+LSTM+T model showed better classification results for four classes, the proposed model closely follows, demonstrating strong training across all 12 classes. 
 
In HCM performances involving Khayal or Dhrupad singing, which constitutes the majority of our dataset, the starting 5 to 10 minutes of the performance generally involves the ``Vilambhit" (very slow tempo) section, where the singer will be singing just straight notes stretched to a few seconds until they start singing in ``Madhya Laya" (medium tempo of the range 100 to 150 BPM). 
Consequently, some of the 30-second audio clips extracted from these early portions contain only a few long, sustained notes. Such clips are likely to challenge the model, as they may appear very similar across different ragas that share similar ``swaras" in their ``Aaroh" (Bihag and Maru-Bihag) and ``Avroh" (Shuddha-Kalyan and Yaman). Despite this inherent difficulty, the model demonstrates remarkable performance.
\subsection{Testing on Other Datasets}
We test our CN2+LSTM+T model on the Saraga dataset\cite{Saraga} and test examples from YouTube. We carefully select audio files from the Saraga dataset corresponding to the 12 Raga classes in our study. We do not find any test files for Nayaki-Kanada in Saraga, so we use test files from our original dataset $S_t$ for this particular class. For the remaining Raga classes, we identify a total of 23 audio files from Saraga, with Bhairavi having the maximum representation of 5 files, while other classes have 3 or fewer, some with only a single file per class. From YouTube, we select 55 files of HCM performances having a minimum of 4 and a maximum of 5 files per Raga class.
We apply all pre-processing as described in Section \ref{sec:Data Pre-processing}. We utilize the tonic values given in the metadata of Saraga dataset.
We test the compiam\cite{compiam_mtg_2023} tool on our dataset and find it to be 91.76\% accurate in predicting tonic values. So, we use this to get tonic values for the YouTube test files.
Our evaluation metrics include both chunk-wise f1-score and song-wise (polling among all the chunks of each song for final prediction) f1-score. 
The comparative results between our test dataset $S_t$, YouTube test data, and Saraga dataset are presented in Table~\ref{tab:Saraga_comparison}.

The performance metrics displayed in Table~\ref{tab:Saraga_comparison} are notably robust, especially considering that the model is not trained on the Saraga dataset or any of the YouTube files, but merely tested on them. These results not only demonstrate the efficacy of our proposed model but also validate the quality and representativeness of our newly curated dataset. The model's ability to generalize well to an external dataset underscores its potential for practical applications in Raga classification tasks.  

\begin{table}[!t]
    \centering
    \caption{\label{tab:Saraga_comparison}Test Results on Saraga and YouTube Test Examples}
    \begin{tabular}{|l|l|l|l|l|l|l|}
    \hline
    \multirow{2}{*}{\textbf{Dataset}} & \multicolumn{3}{c|}{\textbf{Chunk-level}} & \multicolumn{3}{c|}{\textbf{Song-level}}\\
    \cline{2-7}
     & \textbf{Precision} & \textbf{Recall} & \textbf{F1} & \textbf{Precision} & \textbf{Recall} & \textbf{F1} \\
    \hline
    \textbf{$S_t$} & 0.90 & 0.89 & 0.89 & 0.97 & 0.97 & 0.97 \\
    \hline
    \textbf{Saraga} & 0.82 & 0.76 & 0.78 & 0.92 & 0.88 & 0.90\\
    \hline
    \textbf{YouTube} & 0.82 & 0.78 & 0.80 & 0.93 & 0.92 & 0.92\\
    \hline
    \end{tabular}
\end{table}

\begin{table}[!t] 
 \begin{center} 
 \caption{\label{tab:train_whole_dataset}Classification report after training on the whole dataset with 41 distinct Raga classes and 42nd class being the `others' class, which includes all the remaining Raga classes.} 
 \begin{tabular}{|l|l|l|l|} 
  \hline 
  \textbf{True} & \textbf{Precision}& \textbf{Recall}& \textbf{F1-Score} \\ 
  \hline 
  Chunk-Level &0.65 &0.64&0.64  \\ 
  \hline 
  Song-Level &0.80 &0.82&0.81\\
  \hline 
 \end{tabular} 
\end{center} 
\end{table}

\subsection{Train on the whole dataset}
We also train the CN2+LSTM+T model on the whole dataset using the top 42 Raga classes having at least 4 audio files, and the 43rd class being the Others class, consisting of audio files from all the remaining Raga classes. The model demonstrates effective learning on the whole dataset, achieving a chunk-level f1-score of 0.64 and song-level f1-score of 0.80 as shown in Table~\ref{tab:train_whole_dataset}. While these results are not directly relevant to our XAI analysis, we report them to establish a performance benchmark for our dataset when applied across the full range of Raga classes. 
\begin{figure}[!t]
\centering
\includegraphics[width=1\columnwidth]{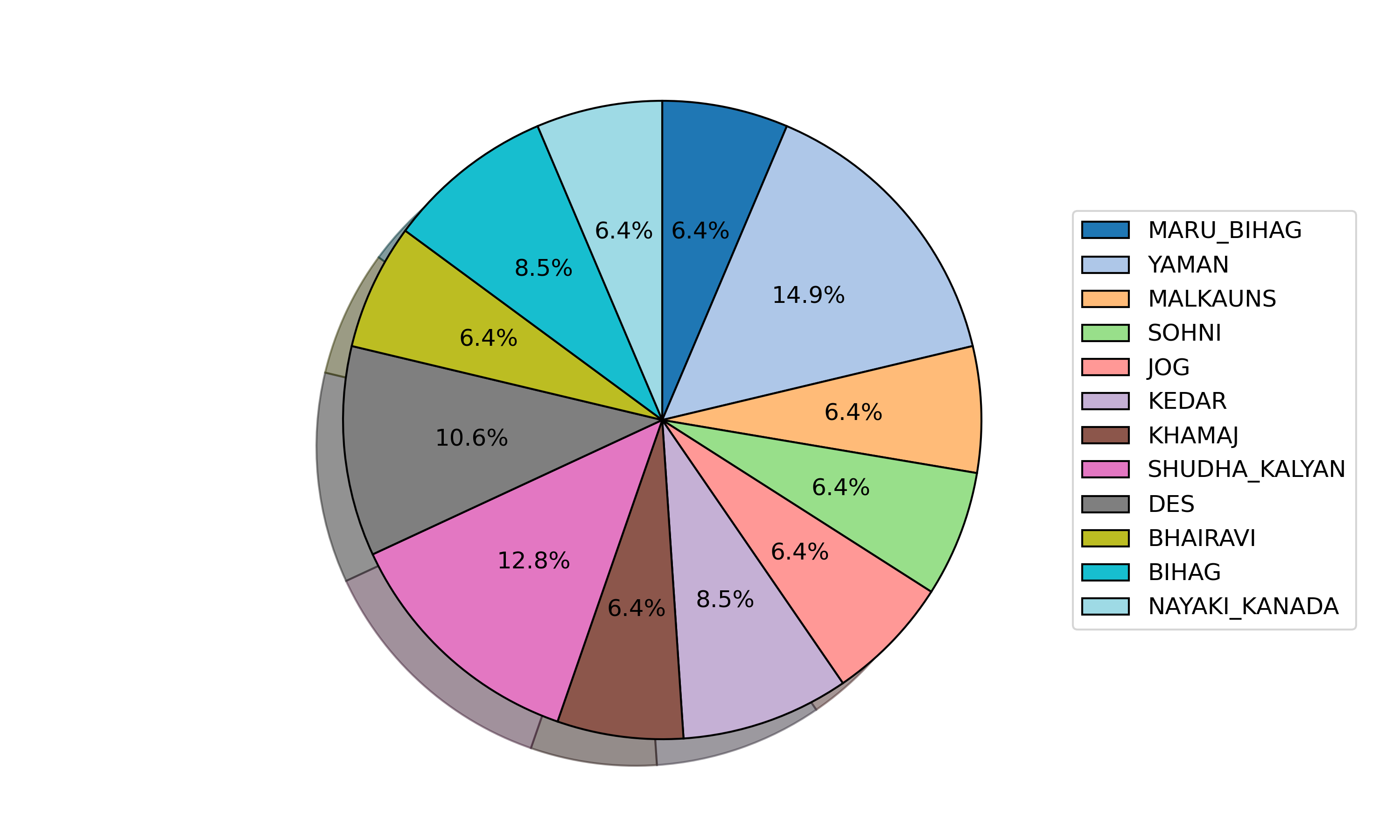}
\caption{Visualisation of Raga class distribution for test files selected for XAI analysis}
\label{fig:Distri}
\end{figure}
\subsection{Saliency Evaluation}
Now, we conduct an in-depth analysis of the salient regions identified by the XAI models to understand their musical significance and validate the manual annotations provided by the music expert. 

\subsubsection{Saliency annotation} 
The music expert is given a randomly selected set of 64 audio chunks out of $S_t$ for which the model predicts the correct class.
The task is to look for regions in each audio file that contain musical elements such as the Pakad, Aaroh, Avroh, or Mukh-Ang of the raga: features that would allow any HCM expert to determine if the audio snippet belongs to a particular raga class. The expert selects only those audio files that have at least 5 to 7 seconds, and at most 15 to 17 seconds of such salient regions as it makes little sense to mark just one or two seconds, or the entire audio, as the salient region. The salient region can be a single segment or may consist of two or three disjoint regions throughout the 30-second audio chunk. We also ensure that we have at least three audio examples per class for all $N=12$ classes. After the annotation process, we obtain 47 labeled test examples $S_{ts} \subset S_t$ to evaluate our saliency methods. The distribution of test classes is shown in Fig.~\ref{fig:Distri}.

\subsubsection{Saliency evaluation}
We compare the salient regions identified by GC and SL with expert annotations to evaluate the models' ability to highlight musically significant regions.
For GC, we sum the activation map values across the frequency axis to obtain a saliency vector over time frames. This vector indicates the importance of each time frame in the model's prediction. For SL, we obtain the coefficients from the linear regression model, representing the importance of each of the 30 time masks. We upsample these coefficients to match the time resolution of the expert annotations.
We calculate the precision of the models' salient regions against the expert annotations for the top N seconds (N = 1 to 10), assessing how well the most critical regions identified by the models align with musically significant parts of the audio.

\begin{table}[!t] 
 \begin{center} 
 \caption{\label{tab:saliency_results_1}Saliency report for GC and SL} 
 \begin{tabular}{|l|l|l|l|} 
  \hline 
  \textbf{Time frame} & \textbf{GC Precision} & \textbf{SL Precision} \\ 
  \hline 
  1 second &0.61 &0.68  \\ 
  \hline 
  2 seconds &0.60 &0.66 \\
  \hline 
  3 seconds &0.59 &0.64 \\
  \hline 
  4 seconds &0.58 &0.60 \\ 
  \hline 
  5 seconds &0.57 &0.58 \\ 
  \hline 
  6 seconds &0.57 &0.58 \\ 
  \hline 
  7 seconds &0.56 &0.58  \\ 
  \hline 
  8 seconds & 0.56 & 0.56  \\ 
  \hline 
  9 seconds &0.55 & 0.55  \\ 
  \hline 
  10 seconds &0.55 & 0.54  \\ 
  \hline 
 \end{tabular} 
\end{center}  
\end{table} 
\begin{figure}[!t]
\centering
\includegraphics[width=1\columnwidth]{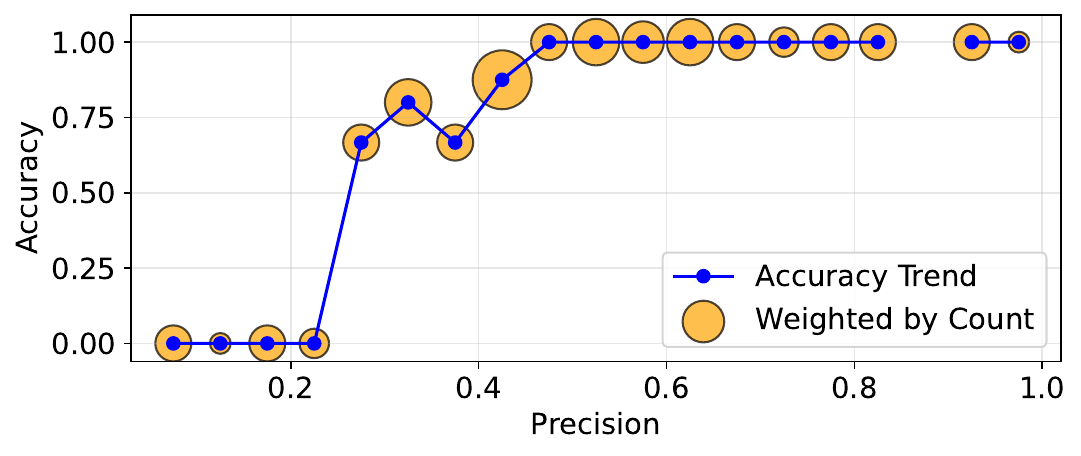}
\caption{Precision vs. Accuracy: The graph shows the relationship between the precision of the XAI model (binned with increments of 0.5) and the classification model's accuracy. The yellow circles represent the weight of each bin, indicating the number of points in each bin.
 }
\label{fig:acc_vs_prec}
\end{figure}
\subsubsection{Saliency results}\label{sec:Saliency results} 
Table~\ref{tab:saliency_results_1} presents the average precision for both the XAI models over the test-set $S_{ts}$, compared with expert annotations.
The precision values are calculated for the top $t$-second ($t\in [1:10]$) regions, where $t=1$ represents how well the most salient 1-second region predicted by the XAI model aligns with the salient region annotated by the expert.
Observe that for a smaller t, precision is high and it decreases as t increases, indicating that the top t seconds of the salient regions highlighted by these XAI models are very close to actual salient regions for that particular audio chunk, implying that the model is indeed relying on the appropriate time frames (the music in those time-frames) to make its predictions for current Raga class.

Let us now observe the relationship between the precision of the XAI model and the accuracy of the classification model. We bin the precision scores of the XAI model with an increment of 0.5. For each bin, we compute the classification model's accuracy and plot Precision vs Accuracy as shown in Fig.~\ref{fig:acc_vs_prec}.
We observe a clear trend showing that higher precision from the XAI model correlates with higher classification accuracy. This indicates that when the XAI model highlights regions with high precision, the classifier is more likely to make correct predictions, indicating that the generated explanations are indeed meaningful and align with human understanding.

From Table~\ref{tab:saliency_results_1}, we observe that SL outperforms GC, particularly for the top few seconds. This behavior is expected, as GC considers activations only from the CNN layers, excluding the contributions of the LSTM layers. While there is no direct metric to compare the two methods, we argue that SL provides more reliable explanations, as it incorporates predictions from the entire model, including the LSTM layers, making it a better representation of the classification model's decision-making process in our case.

\begin{table}[!t] 
 \begin{center} 
 \caption{\label{tab:Analysis of individual samples}Analysis of individual samples} 
 \begin{tabular}{|l|l|l|l|l|} 
  \hline 
  \textbf{True} & \textbf{Predicted}& \textbf{softmax}& \textbf{GC} & \textbf{SL} \\ 
  \textbf{Sample} & \textbf{Sample}& \textbf{probability}& \textbf{Precision} & \textbf{Precision} \\
  \hline 
  Des &Des &0.95&0.77&0.8  \\ 
  \hline 
  Bihag &Maru-Bihag &0.71&0.12&0.0 \\
  \hline 
 \end{tabular} 
\end{center} 
\end{table} 

\begin{figure}[!t]
\centering
\includegraphics[width=1\columnwidth]{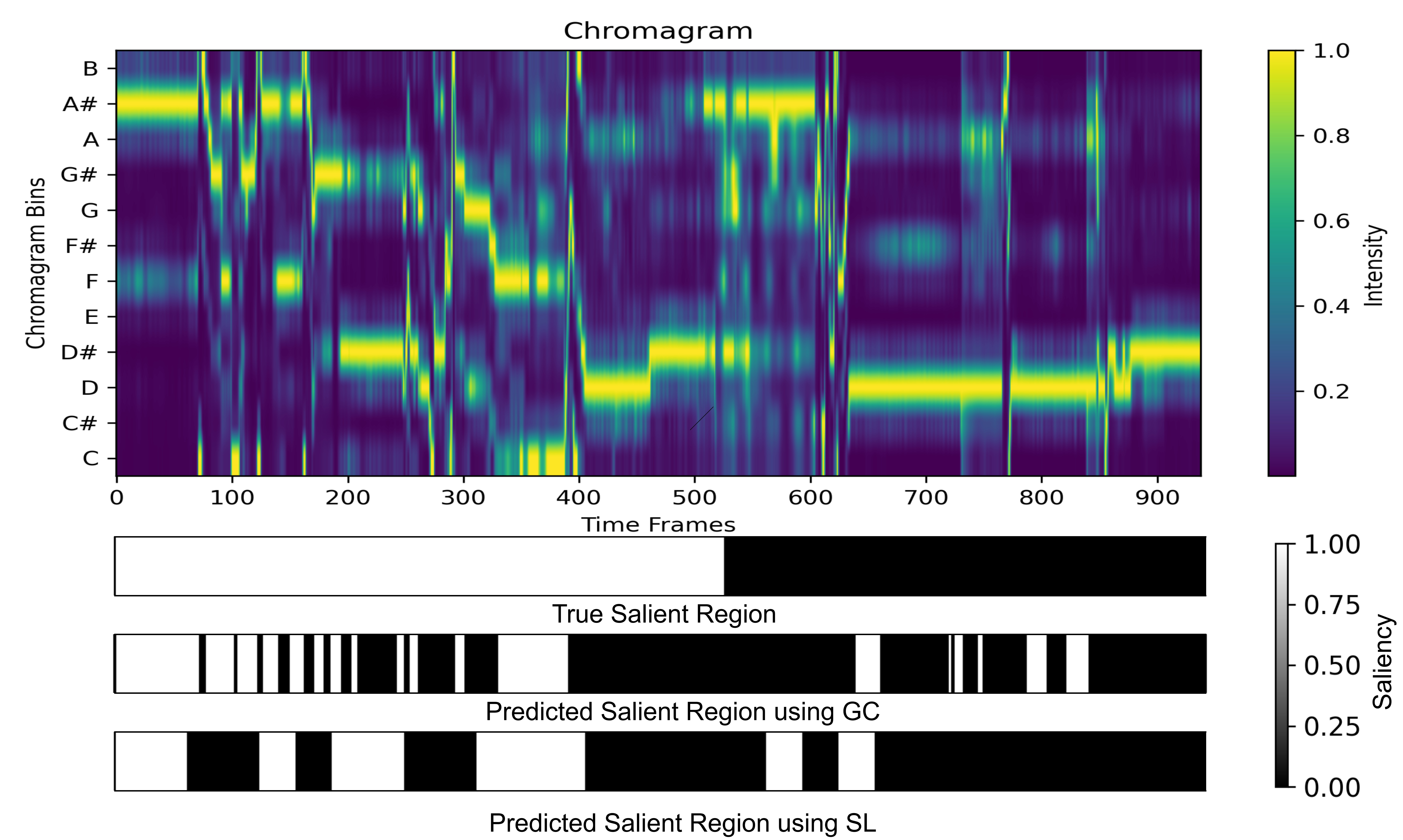}
\caption{Salient Regions for Raga Des sample}
\label{fig:saliency_region_Des}
\end{figure}

\subsection{Musical Analysis of Salient Regions}
\subsubsection{Example 1: Raga Des}
For comparison, we analyze a randomly selected, correctly predicted sample of Raga Des, where the classifier model confidently assigns a softmax probability of 0.95 as shown in Table~\ref{tab:Analysis of individual samples}. For the audio, the expert was able to identify a salient region of about 16 seconds, which includes the note sequence: `\emph{Pa Dha me Ga Re ma Ga Re Ga Ni Sa}', which resembles the pakad (signature phrase) of Raga Des. 
Both GC and SL highlight regions that overlap significantly with this segment annotated by the expert, with precision scores of 0.77 and 0.80 respectively for the top 10 seconds, as shown in Table~\ref{tab:Analysis of individual samples}. 
This significant overlap suggests that the XAI models are looking for musically meaningful regions to make their predictions for salient regions, and are aligning well with human understanding of Raga as a concept.

\begin{figure}[!t]
\centering
\includegraphics[width=1\columnwidth]{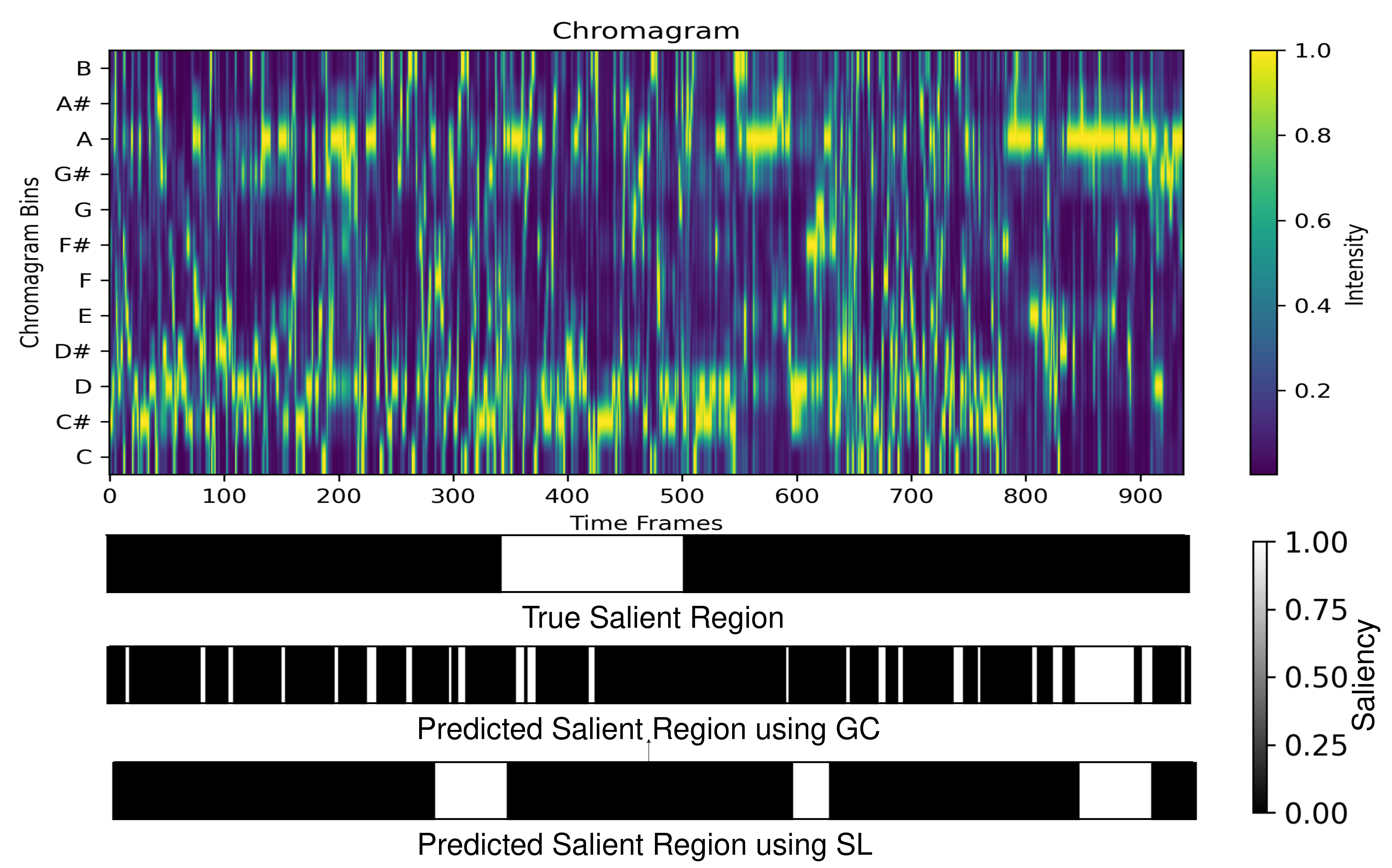}
\caption{Salient regions for Raga Bihag sample}
\label{fig:saliency_region_Bihag}
\end{figure}

\subsubsection{Example 2: Raga Bihag misclassification}
\begin{figure}[!t]
\centering
\includegraphics[width=1\columnwidth]{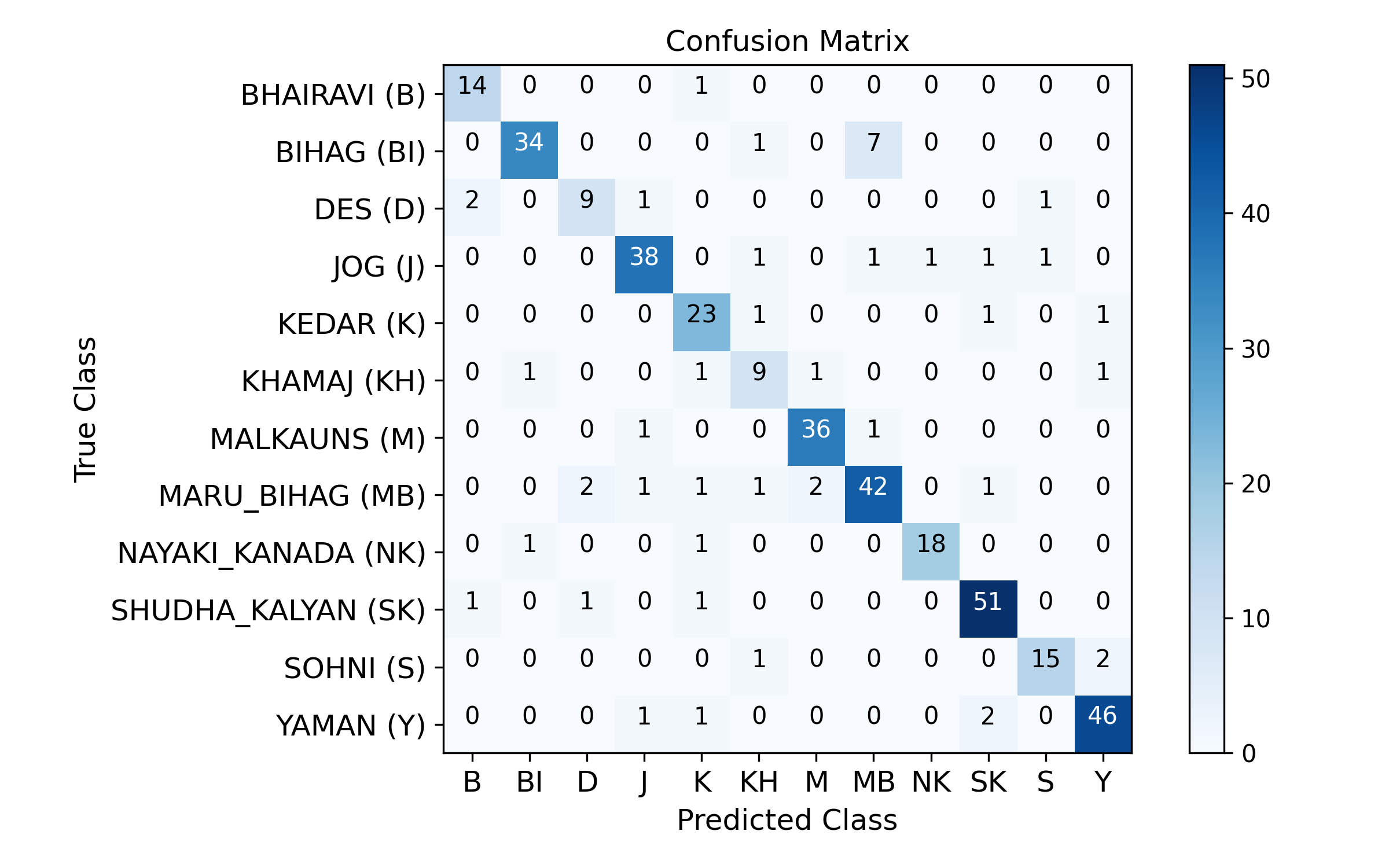}
\caption{Confusion Matrix for predictions on test dataset for CN2+LSTM+T model}
\label{fig:confusion_matrix}
\end{figure}

We now conduct a study on two closely related Ragas in HCM, namely Bihag and Maru-Bihag. Maru-Bihag, being a musical blend of two Ragas, ``Maru" and ``Bihag", is supposed to exhibit some characteristics of Raga Bihag in its composition. 
We can observe in the confusion matrix shown in Fig.~\ref{fig:confusion_matrix} also that there are a few samples of class Bihag that have been misclassified as Maru-Bihag.

We analyze a false negative example where the model incorrectly classifies the true Raga Bihag as Maru-Bihag. The annotated salient region for Bihag, along with the predicted regions from GC and SL, is shown in Figure~\ref{fig:saliency_region_Bihag}.
The softmax probabilities (Table~\ref{tab:Analysis of individual samples}) indicate 0.26 for the true class (Bihag) and 0.71 for the predicted class (Maru-Bihag), suggesting that the model partially recognizes the true class but is misled by overlapping characteristics between the two Ragas present in the test audio.

Expert annotations reveal that only a small portion of the audio distinctly represents Raga Bihag (Fig.~\ref{fig:saliency_region_Bihag}), indicating inherent ambiguity in the audio itself. This is further reflected in the low precision scores for GC (0.12) and SL (0.0) in Table~\ref{tab:Analysis of individual samples}, as their predicted salient regions fail to align with the annotated region and instead highlight unrelated sections of the audio. The mismatch suggests that the audio contains regions with characteristics closer to Maru-Bihag, while the distinctive features of Bihag are less prominent. This observation underscores the complexity of distinguishing between closely related Ragas and is corroborated by expert listening and annotation.

\section{Conclusion and Future Scope}\label{sec:conclusion_and_future_scope}
In this paper, we curate a
dataset for Hindustani Classical Music analysis from Prasar-Bharti audios, manually annotate it with Raga and Tonic labels, and achieve an f1-score of 0.89 for the 12-class Raga classification task using the proposed model. 
We observe that providing tonic normalization to the chromagram features significantly enhances the performance of the Raga classifier. 
Following that, we employ two Explainable AI (XAI) models: GradCAM++ and SoundLIME, to identify the salient regions for the classifier's predictions and analyze their overlap with human annotations. 
Our comparison reveals a significant overlap, affirming that the patterns or characteristics theoretically associated with the identification of a Raga for any audio are present and are being used by the
Automatic Raga Identification (ARI) model for the task. 
This serves as a correctness check for the music theory that specific motifs, phrases or characteristics are indeed responsible for the identification of any Raga 
by the Deep Learning model built on a dataset that represents actual HCM performances.
It is important to note that experts may not always agree on the same set of phrases or regions, indicating the subjective nature of such music analysis. 

This work provides a valuable contribution to the field of explainability in ARI for Indian Art Music, setting the foundation for future research. 
Moving forward, incorporating musicians' perspectives and musical concepts like ``Khatka" and ``murki" to refine Raga classifiers can ensure correctness of ARI according to musical conventions and nuances. 
Future research can focus on developing methods to generate listenable or musical explanations, which could serve as standalone identifiers of Raga. This can include extracting phrases from audio clips that correspond to the Raga, offering potential applications in music pedagogy and beyond.

\bibliographystyle{IEEEtran}
\bibliography{bib}

\vfill

\end{document}